\documentclass{article}

\usepackage{arxiv}

\usepackage[utf8]{inputenc} 
\usepackage[T1]{fontenc}    
\usepackage{url}            
\usepackage{booktabs}       
\usepackage{amsfonts}       
\usepackage{nicefrac}       
\usepackage{microtype}      
\usepackage{lipsum}		
\usepackage{graphicx}

\usepackage{cite}
\usepackage{amsmath,amssymb,amsfonts}
\usepackage{algorithmic}
\usepackage{graphicx}
\usepackage{textcomp}
\usepackage{xcolor}
\usepackage{array}
\usepackage{ marvosym }

\usepackage{bm}
\usepackage{float}
\usepackage{mathtools}
\usepackage{ mathrsfs }
\usepackage{ amssymb }
\usepackage{kantlipsum}
\usepackage{ dsfont }
\usepackage[ruled,vlined]{algorithm2e}

\title{
    Adversarial Filters for Secure Modulation Classification
}

\author{
    Alex Berian, Kory Staab, Noel Teku, Gregory Ditzler, Tamal Bose, Ravi Tandon \\ Department of Electrical and Computer Engineering \thanks{ This project was partially supported by the Department of Energy (DOE) \#DE-NA0003946. This project was also partially supported by the Broadband Wireless Access and Applications Center (BWAC); NSF Award No. 1822071 }\\  The University of Arizona\\ Tucson, Arizona, USA \\ \{berian, kstaab, nteku1, ditzler, tbose, tandonr\}@email.arizona.edu
}

\date{}

\begin{document}
\maketitle

\begin{abstract}
Modulation Classification (MC) refers to the problem of classifying the modulation class of a wireless signal. In the wireless communications pipeline, MC is the first operation performed on the received signal and is critical for reliable decoding. This paper considers the problem of secure modulation classification, where a transmitter (Alice) wants to maximize MC accuracy at a legitimate receiver (Bob) while minimizing MC accuracy at an eavesdropper (Eve).

The contribution of this work is to design novel adversarial learning techniques for secure MC. In particular, we present adversarial filtering based algorithms for secure MC, in which Alice uses a carefully designed adversarial filter to mask the transmitted signal, that can maximize MC accuracy at Bob while minimizing MC accuracy at Eve. We present two filtering based algorithms, namely gradient ascent filter (GAF), and a fast gradient filter method (FGFM), with varying levels of complexity.

Our proposed adversarial filtering based approaches significantly outperform additive adversarial perturbations (used in the traditional ML community and other prior works on secure MC) and also have several other desirable properties.  In particular, GAF and FGFM algorithms are a) computational efficient (allow fast decoding at Bob), b) power-efficient (do not require excessive transmit power at Alice); and c) SNR efficient (i.e., perform well even at low SNR values at Bob).
\end{abstract}

\section{Introduction}

In recent years, ML has had success in many applications such as speech recognition \cite{speech}, stock prediction \cite{stock}, and natural language processing \cite{nlp}. Most notably, deep learning (DL) has surpassed human ability in image classification \cite{human_bad}.

In the realm of wireless communication ML has recently been adapted for many uses such as equalization \cite{equalization}, spectrum sensing \cite{spectrum_sensing}, channel coding \cite{channel_coding}, signal classification \cite{oshea_iq_original}, etc.
One important use of ML in wireless communication is modulation classification (MC). In ML classification problems, ML models take an input and associate it with a class. In MC, the input to a ML model is a set of features extracted from a signal or samples of the signal itself, and the classes are the possible modulation formats used to generate the signal.

There are two categories of MC: likelihood-based methods, and feature extraction approaches.
Likelihood-based methods seek to calculate the likelihood of a signal belonging to a certain class. While optimal from a statistical perspective, these approaches can be too computationally complex in most applications.  \cite{garrett_hierarchical}.
Feature extraction involves prepossessing the signal to gather statistics that can be used as features to classify the modulation format. Many works have shown that extracting features (e.g., cyclostationary features, higher-order statistics, etc.) can be reliable inputs for ML MC models \cite{mc_survey}.
Instead of relying on crafted features for classification, one can pass the raw I/Q samples of the signal into a neural network (NN) which internally develops features through nonlinear transformations \cite{oshea_iq_original}. This is beneficial because features developed by NNs are optimized as opposed to human-constructed features, which are often selected based on empirical evidence.
Although NNs on their own are quite successful in MC, it has been shown that combining a NN acting on raw IQ data with a feature extraction classification approach improves performs better than the two approaches individually \cite{combined_mc}.
The ability to classify modulation formats in real-time allows for blind modulation to improve data rate, where a transmitter does not communicate to a receiver what the modulation format is, so the receiver must classify it then decode data. \cite{mc_survey}.

Despite the benefits of using ML for classification, adversarial learning (AL) shows that these algorithms are easily fooled by \emph{adversarial examples}, which are created by strategically crafting changes to inputs \cite{first_al_paper}.
Many existing input transformation AL attacks involve adding a small perturbation to the input. Different algorithms such as the fast gradient sign method (FGSM) \cite{goodfellow_adversarial} and the Jacobian-based saliency map \cite{jbsm} attack use the gradient of the loss with respect to the input to the ML to find small perturbations to the input that cause misclassification. Iterative approaches for creating adversarial examples exist as well, such as deep fool \cite{deepfool}. Some of the aforementioned gradient-based attacks are input specific, meaning the crafted perturbations are designed for a particular input to a particular fixed ML model. Universal adversarial perturbations (UAP) are perturbations that work for any input for a fixed ML model \cite{uap}.

It has also been shown that AL attacks can have detrimental effects on MC \cite{sadeghi_original}. It has been shown that a low power jammer sending adversarial perturbations can significantly impede devices trying to classify modulation formats of signals \cite{al_mc_jam}.
In addition to blind modulation, MC also has application in other military and commercial scenarios. One such military scenario is classifying the modulation format of detected signals to determine if it is from friend or foe, then use triangulation to locate them. Primary users in licensed spectrum can use MC to identify users that are not adhering to protocols of the spectrum.
AL is one method transmitters can prevent others from classifying their modulation format. This will make it difficult for enemies in military situations to identify a transmitter as an enemy transmitter.

This paper analyses an eavesdropper scenario where a transmitter (Alice), has an intended receiver (Bob), and there is an eavesdropper (Eve) listening in on the transmission. The communication system uses blind modulation, so both Bob and Eve need to classify the modulation format before they can begin to decode information. There exist applications where decoding is not desired, simply classifying the modulation format is sufficient for identification of other parties.
Encrypting the data before transmission does not impair Eve's ability to classify the modulation format a signal, so it is necessary to use physical layer approaches to prevent her ability to classify.
Before sending an intended signal, Alice uses AL to send adversarial examples instead.
Alice can share a key with Bob offline to help Bob undo the adversarial attack, classify the signal, and ultimately decode information. Eve will not have this key making it difficult to classify the modulation format.
For a transmitter to fool an eavesdropper's classifier and allow for intended receivers to undo the attack, a UAP must be used. Alice cannot construct a new perturbation for every signal sent, because Bob cannot know every possible perturbation and when they are used.
We measure secrecy in the sense that Bob's classification accuracy remains high, but Eve's classification accuracy is low.

There are trade-offs to consider when Alice has limited transmit power and the received signal at Bob must satisfy a minimum signal to noise ratio (SNR) constraint to ensure decodability.
Most AL algorithms for generating adversarial examples involve adding a perturbation to the input. When Alice sends an adversarial example generated by adding a perturbation, she must allocate some of her finite transmission power to the perturbation.
In doing so the SNR at Bob will decrease, so to reduce Eve's classification accuracy as much as possible Alice must allocate just enough power to the intended signal to satisfy the SNR constraint and allocate the rest of the transmission power to the additive perturbation. Such additive perturbation strategies may not be feasible with limited transmit power at Alice.
Typically MC models only observe a finite amount of samples, so the additive perturbation must also be the same number of samples. Real-time communication lasts many more samples than just the input size to the MC model, so the additive perturbation can simply be repeated. Although a receiver with a finite input MC model may not observe a window where the perturbation is aligned, it has been shown that such perturbations are shift-invariant and will still reduce classification accuracy \cite{sadeghi_original}.
If Bob knows the perturbation being used, he can subtract it off then classify and decode. To subtract away the perturbation, Bob needs to sync his subtraction with Alice's addition, because Alice is continuously repeating the perturbation.
Bob must also make sure that he scales the perturbation before subtracting to ensure that there is not any more trace of the perturbation in the signal, since the transmitted signal from Alice experiences attenuation.

In this paper, we show that it is also possible to use a filter to create adversarial examples instead of an additive perturbation. Filter-based adversarial attacks do not suffer from the aforementioned drawbacks in this eavesdropper scenario (i.e., alignment searching, attenuation prediction, power limitations).
When Alice uses a digital filter to create the adversarial example that is transmitted then Bob can use the inverse of the filter to get the original signal back. There is no need to sync or find the attenuation like with additive perturbations.
Additive perturbations require that Alice sacrifice some transmission power. Digital filters do not require sacrificing transmit power. Instead, power within the signal is being transferred between different frequencies, so there is no need to sacrifice transmission power to create effective adversarial examples.
Not sacrificing transmission power also implies that the SNR at Bob will increase when the filter is used over an additive perturbation.

The main contributions of this paper are the following:
Two novel methods of creating adversarial examples using filters are proposed. The first approach is an iterative optimization technique where filter taps are effectively "trained". The second is an analytical solution to the optimization problem of maximizing loss with respect to the filter's taps.
The proposed approaches are for creating finite impulse response (FIR) filters, where the inverse filters are infinite impulse response (IIR) filters that are trivial to solve for.
This paper analyses an eavesdropper scenario for MC, where secrecy is measured by the reduction in classification accuracy at the eavesdropper by applying adversarial attacks at the transmitter. Benefits and drawbacks of using traditional additive perturbations versus the proposed filter methods are employed are analyzed over the metrics of classification accuracy at the eavesdropper, available transmit power at the transmitter, and SNR requirement at the intended receiver.

The organization of the remainder of the paper is as follows. Section \ref{sec:sys_model} states the system model of the problem trying to be solved. \ref{sec:results} discusses the important findings of this paper with regards to the system model. \ref{sec:filter_algs} presents novel filter-based AL algorithms. In section \ref{sec:experiments} simulation experimental results are presented and analyzed. \ref{sec:conc} concludes.

\section{System Model \& Problem Statement}

\label{sec:sys_model}
\begin{figure*}[t]
    \centering
    \includegraphics[trim=0 0 357 165, clip,angle=-90,width=\textwidth]{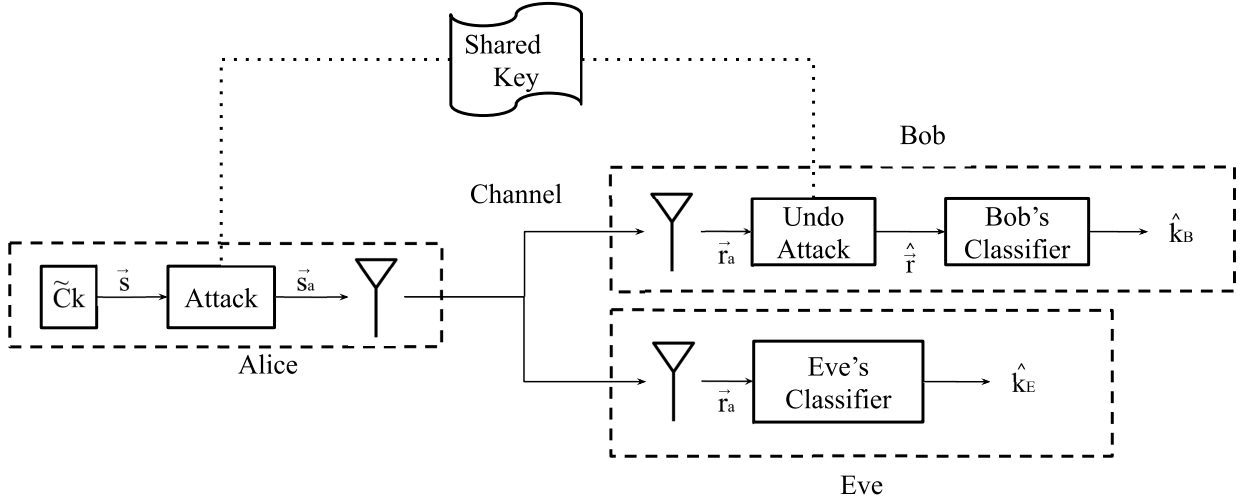}
    \caption{Assumed Communication system model with Transmitter ({Alice}), receiver ({Bob}), and eavesdropper ({Eve})}
    \label{fig:mod_system}
\end{figure*}

The system model is displayed in Figure \ref{fig:mod_system}.
Alice wants to send a signal $\vec{s}$ to Bob which is a $[d \times 1]$ vector of complex values ($\vec{s} \in \mathds{C}^d$).
There are different ways of structuring binary data into a signal $\vec{s}$ that can be sent over a communication link. The different structures are the possible $K$ modulation formats with which Alice transmits information to Bob.
$\vec{s}$ is a signal with modulation format $k$, and $\tilde{C}_k$ is the set of all signals from class $k$ (i.e $\vec{s} \in \tilde{C}_k$).
Alice wants to keep $k$ secure from Eve. Alice may change the modulation format $k$ for different blocks of transmitted data depending on channel conditions or other motivations.
Using AL, Alice creates an adversarial example $\vec{s}_a$ from $\vec{s}$. Using some adversarial attack function $f$ with an $[m\times 1]$ vector of parameters $\vec{\delta}$,
\begin{equation}
    \label{adversarial_attack}
    \vec{s}_a = f(\vec{s},\vec{\delta}).
\end{equation}
Alice transmits $\vec{s}_a$ and has a finite transmission power $P_T$, this introduces the constraint
\begin{equation}
    \|\vec{s}_a\|^2 \leq P_T,
\end{equation}
Where $\|\cdot\|$ is the L2 norm operation.
The communication channel from Alice to Bob and Eve are assumed to be identical, and the input to the channel $\vec{s}_a$ has the output $\vec{r}_a$, and the input $\vec{s}$ would have the output $\vec{r}$. The channel is comprised of attenuation of the transmitted signal, and additive noise: 
\begin{equation}
    \vec{r}_a = \alpha \vec{s}_a + \vec{N},
\end{equation}
where $\alpha$ is a real scalar attenuation coefficient, and $\vec{N}$ is a complex additive white gaussian noise (AWGN) vector with zero mean.
Alice and Bob have a pre-shared key. In this problem, the key is assumed to already have been securely shared between both Alice and Bob with no overhead.
Bob uses this key to try to undo the adversarial attack to get $\hat{\vec{r}}$ which is his on $\vec{r}$. Bob the passes $\hat{\vec{r}}$ into his classifier $h_{B}$ to get his guess $\hat{k}_B$ of true modulation format $k$:
\begin{equation}
    \hat{k}_B = h_B(\hat{\vec{r}}).
\end{equation}
Once Bob chooses the modulation format $\hat{k}_B$, he can use the appropriate demodulation scheme. For communication to be reliable, $\hat{\vec{r}}$ must satisfy a minimum SNR.
Eve does not have this key and can't find $k$ from the received signal $\vec{r}_a$ using her classifier $h_E$.
Eve's classification accuracy is the probability that her classifier acting on the received signal outputs the correct class:

\begin{equation}
    P_E = Pr\{h_E(\vec{r}_a) = k\} = Pr\{\hat{k}_E = k\}.
\end{equation}
Bob's classification accuracy is the probability that his classifier acting on the recovered received signal outputs the correct class.
\begin{equation}
    P_B = Pr\{h_B(\hat{\vec{r}}) = k\} = Pr\{\hat{k}_B = k\}.
\end{equation}
Alice and Bob's goal is to design an adversarial attack and shared key that keeps Eve's classification accuracy $P_E$ low and Bob's classification accuracy $P_E$ high, while satisfying the constraint that the SNR of recovered signal $\hat{\vec{r}}$ remain above a constant threshold. The minimum SNR at Bob is application dependent.
This paper analyses the trade-offs between transmit power $P_T$, the SNR requirement at Bob, and Eve's classification accuracy when Alice has full control over $f$, $\vec{\delta}$, and the shared key is $\vec{\delta}$.

For additive attacks, the attack function in \ref{adversarial_attack} is defined as
\begin{equation}
    f_{+}(\vec{s},\vec{\delta}_+) = \vec{s} + \vec{\delta}_+,
\end{equation}
and $m=d$ to satisfy dimensionality constraints. $\vec{\delta}_+$ is the additive perturbation in additive attacks.
In a filter-based attack, (\ref{adversarial_attack}) can be rewritten as follows:
\begin{equation}
    f_{\circledast}(\vec{s},\vec{\delta}_{\circledast}) = \vec{s} \circledast \vec{\delta}_{\circledast},
\end{equation}
and $m \neq d$. $\vec{\delta}_{\circledast}$ denotes the taps of the FIR filter that generates the adversarial sample, and $\circledast$ is convolution between two finite vectors.
This work only analyses the performance of additive and filter-based forms of the adversarial attack $f(\vec{s},\vec{\delta})$.

To correctly classify the received signal and decode, Bob must reverse the adversarial attack from $\vec{r_a}$ to get $\hat{\vec{r}}$. That is
\begin{equation}
    \hat{\vec{r}} = f^{-1}(\vec{r_a},\vec{\delta}).
\end{equation}
The formulation of $f^{-1}(\vec{r}_a,\vec{\delta})$ depends on how the key is shared between Alice and Bob. In this paper the shared key is assumed to be the vector $\vec{\delta}$. $\vec{\delta}$ is assumed to be a secret pre-shared key between Alice and Bob with no communication overhead. The key is never altered, nor shared during communication between Alice and Bob. The function $f(\cdot,\cdot)$ is known by both Bob and Eve.

\section{Main Results \& Discussion}
\label{sec:results}

\subsection{Reversing Adversarial Attacks at Bob}
Bob needs to know how the adversarial attack is performed to effectively undo the adversarial attack. Bob needs to know the formulation of $f(\vec{s},\vec{\delta})$, as well as the vector $\vec{\delta}$. The proposed shared key between Alice and Bob is the function $f(\cdot,\cdot)$ and the vector of parameters $\vec{\delta}$.

Reversing additive attacks requires subtracting $\vec{\delta}_+$ from $\vec{r}_a$ with the proper attenuation.
\begin{equation}
  \hat{\vec{r}}_+ = \vec{r}_a - \alpha \vec{\delta}_+ = \alpha \vec{s} + \vec{N}
  .
\end{equation}
However, Bob cannot perfectly find $\alpha$, so he must make an estimate $\hat{\alpha}$ of the true $\alpha$. This fault alters the expression of $\hat{\vec{r}}_+$ to
\begin{equation}
    \label{eq:alpha_hat}
    \hat{\vec{r}}_+ = \alpha \vec{s} + \vec{N} + (\alpha - \hat{\alpha}) \vec{\delta}_+
    .
\end{equation}
From (\ref{eq:alpha_hat}) it is obvious that unless Bob perfectly estimated $\vec{\alpha}$ there will always be remnants of the adversarial perturbation $\vec{\delta}_+$. This is particularly bad for Bob because many AL algorithms that generate $\vec{\delta}_+$ are designed to create perturbations that ruin classification accuracy even when the perturbation's power is extremely small \cite{first_al_paper}\cite{goodfellow_adversarial}\cite{cw}. Additive perturbations are comprised of a finite number of samples, but real-time implementation consists of a never-ending stream of samples. In real-time Alice can immediately retransmit the finite perturbation whenever she finishes transmitting its finite number of samples. Even if Eve's classifier looks at a window where the perturbation is not perfectly lined up, her classification accuracy is still low due to the shift-invariance property of additive adversarial attacks for MC \cite{sadeghi_original}. For Bob to undo this attack he must sync his removal of $\vec{\delta}_+$ with Alice's repeated transmission. This is a one time process during communication. However, it requires more complex hardware to implement than if a filter was used.

Reversing a filter attack is a much simpler process. This paper focuses on FIR filters, with an IIR inverse filter. Let $\vec{\delta}_{\circledast i}$ be the $i^{th}$ element of $\vec{\delta}_{\circledast}$ (i.e., the $i^{th}$ filter tap). The recursive difference equation for the IIR inverse filter is
\begin{equation}
    \label{eq:inv_filt}
    y(n) = \frac{1}{\vec{\delta}_{\circledast 0}}\bigg(x(n) - \sum_{i=1}^{m-1}\vec{\delta}_{\circledast i}y(n-i)\bigg).
\end{equation}
Let $\delta_{\circledast}^{-1}(j)$ be the impulse response of the inverse filter. The recovered signal under a filter-based attack $\hat{\vec{r}}_{\circledast}$ is
\begin{equation}
    \hat{\vec{r}}_{\circledast} = \vec{\delta}_{\circledast}^{-1} \ast \vec{r}_a = \vec{s} + \vec{\delta}_{\circledast}^{-1} \ast \vec{N}.
\end{equation}
Therefore, when Bob undoes the filter attack, he colors the noise.

\begin{figure*}[t]
    \centering
    \includegraphics[trim=20 50 360 210, clip,angle=-90,width=.7\textwidth]{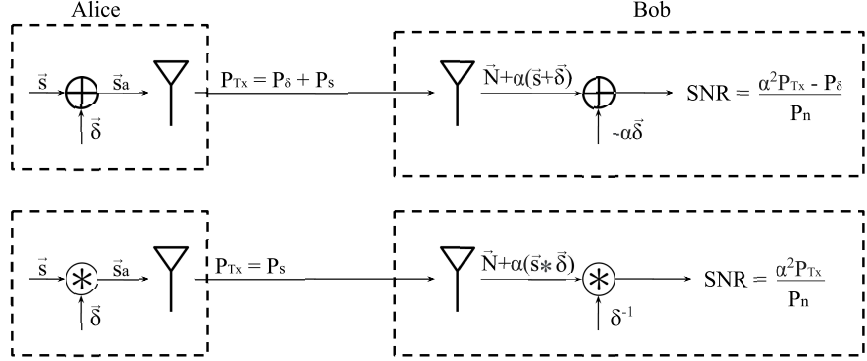}
    \caption{Illustration of how Alice and Bob use AL learning in their communication system. We see that there is a loss in SNR at Bob when an additive perturbation is employed. When a filter is used, there is no loss in SNR.}
    \label{fig:filter_benefit}
\end{figure*}
Figure \ref{fig:filter_benefit} shows the SNR of the recovered signals at Bob when additive and filter attacks are used. Alice has limited transmission power $P_T$. The recovered signal's SNR can be expressed as a function of transmit power $P_T$, perturbation power $P_{\vec{\delta}}$, the attenuation coefficient $\alpha$, and the noise power $P_{\vec{N}}$. Note that $P_{\vec{\delta}}$ is only applicable if an additive perturbation is used.

In the case of additive perturbations, the $SNR_+$ is dependent on the power allocated to the perturbation $P_{\vec{\delta}}$. The transmit power $P_T = P_{\vec{s}} + P_{\vec{\delta}}$ must be allocated to the perturbation $\vec{\delta}_+$ and the intended signal $\vec{s}$, where $P_{\vec{s}}$ is the signal power. The best case scenario for Bob with additive perturbations is that his prediction of the attenuation coefficient is perfect (i.e $\hat{\alpha} = \alpha$). Assuming Bob perfectly eliminates the additive perturbation, the of SNR of the recovered signal $\hat{\vec{r}}_+$, expressed in (\ref{eq:alpha_hat}), for additive perturbations is
\begin{equation}
    SNR_+ = \frac{\alpha P_T-P_{\vec{\delta}}}{P_{\vec{N}}}.
\end{equation}

The FIR filter coloring the noise does not impact the noise power $P_{\vec{N}}$. As for filter-based adversarial attacks used by Alice there is no need to allocate transmit power to a perturbation, so the SNR at Bob $SNR_{\circledast}$ for filter attacks is
\begin{equation}
    SNR_{\circledast} = \frac{\alpha P_T}{P_{\vec{N}}}.
\end{equation}

With a fixed transmit power $P_T$, it is clear that there is an SNR gain to using filter-based attacks over additive attacks. Alice need not exceed the SNR requirement at Bob. Alice can allocate just enough power to $P_{\vec{s}}$ for the requirement, and allocate the rest of $P_T$ to $P_{\vec{\delta}}$ to minimize Eve's classification accuracy. If there is barely enough transmit power to satisfy the SNR requirement, little to no power can be allocated to the perturbation. This results in high classification accuracy for Eve. If a filter-based attack is used, Alice then can still use an effective adversarial attack when there is no power available to allocate to an additive perturbation.

Bob needs to reverse the adversarial attack. Alice cannot pre-share every possible $\vec{\delta}_j$ for every possible signal $\vec{s}_j$ she sends.
Many AL algorithms are designed to create an attack designed for a specific input, such as the FGSM.
Further, AL algorithms that are inherently designed to develop universal attacks, such as the deep fool (DF) algorithm. A universal adversarial attack is one designed to work on any input.
There exist algorithms that take many input specific additive perturbations, and create a universal adversarial perturbation (UAP) \cite{uap}\cite{sadeghi_original}.
Sadeghi et al present an algorithm for generating UAPs with principal component analysis (PCA \cite{pca}) which works well on MC when used on input specific perturbations generated by the fast gradient method (FGM) \cite{sadeghi_original}. The algorithm is to create a vector in the direction of the first principal component of all the $\vec{\delta}_j$.
In addition to attacks being universal, they can also be class-specific. Instead of $\vec{\delta}$ being designed to work on every input $\vec{s}_j \in \tilde{C}$, $\vec{\delta}^{(k)}$ can be designed to work especially well for inputs from the $k^{th}$ class $\vec{s}_j \in \tilde{C}_k$. There are only a finite number of classes, so it is feasible for Alice to pre-share every $\vec{\delta}^{(k)}$ with Bob.

One important aspect regarding undoing the filter at Bob is the invertibility of the filter at Alice.
If the inverse filter at Bob $\vec{\delta}_\circledast^{-1}$, expressed in (\ref{eq:inv_filt}) is unstable, the noise at Bob's receiver will be amplified too much and no communication will be possible.
The inverse filter $\vec{\delta}_\circledast^{-1}$ may not be stable if $\vec{\delta}_\circledast$ does not satisfy certain constraints.
The zeros of $\vec{\delta}_\circledast$ will become the poles of $\vec{\delta}_\circledast^{-1}$. Any filter with poles inside the unit circle is stable. To ensure $\vec{\delta}_\circledast^{-1}$ is stable, $\vec{\delta}_\circledast$ must be designed such that its zeros are inside the unit circle. An FIR filter with all of its zeros inside the unit circle is known as a \emph{minimum phase} filter, which will in turn have a stable inverse.
Alice needs to employ a minimum phase filter so that Bob's inverse filter is stable.

The filters used at both Bob and Alice must have an overall gain of 1 to ensure that signal power is preserved.
\begin{equation}
    \label{eq:power_preserve_freq}
    \frac{1}{2\pi}\int_{-\pi}^{\pi}|D_\circledast(\omega)|^2 = 1,
\end{equation}
where $D_\circledast(\omega)$ is the discrete-time-fourier-transform (DTFT) of the filter $\vec{\delta}_\circledast$. Using Parseval's theorem, the constraint in (\ref{eq:power_preserve_freq}) can be rewritten as the L2 norm of the magnitude: 
\begin{equation}
    \label{eq:power_preserve}
    \||\vec{\delta}_{\circledast i}|\|^2 = 1.
\end{equation}
Here the magnitude operation $|\cdot|$ is element-wise for the complex vector within.
This is important, because Bob does not want to amplify noise power, and Alice does not want to underutilize or surpass her transmit power.

\section{Adversarial Filtering Algorithms}
\label{sec:filter_algs}

\subsection{Gradient Ascent Filter}

\begin{figure}
    \centering
    \includegraphics[trim=30 170 431 375, angle=-90,clip,width=250pt]{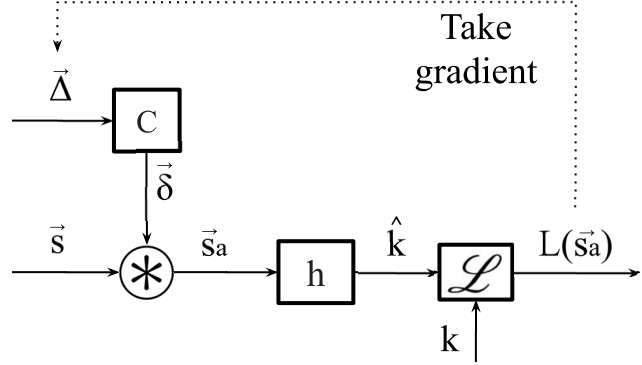}
    \caption{Flowgraph for Training Gradient Ascent Filter (GAF). }
    \label{fig:grad_ascent}
\end{figure}
The first novel filter-based AL algorithm is the gradient ascent filter (GAF). It is an optimization approach to generating an adversarial filter.
A temporary vector $\vec{\Delta}$ of size $[l \times 1]$ is randomly initialized, then converted into a the adversarial filter $\vec{\delta}$ of size $[m \times 1]$ using some filter creation function:
\begin{equation}
    \label{eq:creation}
    \vec{\delta} = C(\vec{\Delta}).
\end{equation}
The desired number of filter taps $m$ and the creation function $C$ determine the size of the temporary vector $l$. The filter $\vec{\delta}$ is a complex valued vector for this MC problem, and depending on the filter creation function $C$, the temporary vector $\vec{\Delta}$ can be real or complex valued.




During the optimization process, the temporary vector $\vec{\Delta}$ is updated iteratively to increase loss as shown in Figure \ref{fig:grad_ascent}. For $\vec{\Delta}$ to be trainable, the filter creation function $C$ needs to be differentiable.
The temporary vector $\vec{\Delta}$ is trained with mini-batch stochastic gradient ascent (SGD), or another optimization process as desired (i.e., Adagrad, RMS prop, Adam \cite{adam}).
The training data for creating the GAF can be any desired subset of all training data, so one can create a universal attack, class-specific attack, or an input specific attack.

The formal process of the GAF algorithm is expressed in algorithm \ref{alg:gaf}, but explained in this paragraph. The algorithm requires Eve's classifier (fixed), the desired number of filter taps $m$, the filter creation function $C$, the loss function to maximize $\mathscr{L}$, set of desired training data signals $\bm{s}$ and it's set of true labels $\bm{k}$, the learn rate for SGD $\eta$, and the desired number of training epochs $T$. The first temporary vector $\vec{\Delta}_0$ is initialized randomly  and i.i.d. as an $[l \times 1]$ Gaussian random vector with zero mean and unit variance. Then the training loop begins with iteration counter $t$ going from $1$ to $T$. In the training loop at iteration $t$, $\vec{\Delta}_{t-1}$ is to create the adversarial filter for this training step $\vec{\delta} = C(\vec{\Delta})$. $\vec{\delta}$ is convolved with all signals in $\bm{s}$ to create the adversarial examples, then the classifier $h$ acts on those examples to create the set of prediction $\hat{\bm{k}}$. The loss for this training step $L$ is calculated as the sum of the loss function acting on every prediction in the set $\hat{\bm{k}}$ and their corresponding true labels in the set $\bm{k}$. The temporary vector at this training step $\vec{\Delta}_{t}$ is updated as the addition of the previous step's temporary vector $\vec{\Delta}_{t-1}$ ,and the gradient of the of the loss with respect to the previous temporary vector scaled by the learn rate $\vec{\Delta}_{t}=\vec{\Delta}_{t-1}+\eta \nabla_{\vec{\Delta}_{t-1}}L$. When the training loop is complete the final temporary vector $\vec{\Delta}_T$ converted into the final adversarial filter $\vec{\delta}_{GAF} = C(\vec{\Delta}_T)$.

\begin{algorithm}
 \caption{Algorithm for creating $\vec{\Delta}$ using the GAF technique}
 \label{alg:gaf}
 \begin{algorithmic}[1]

 \renewcommand{\algorithmicrequire}{\textbf{Inputs:}}
 \renewcommand{\algorithmicensure}{\textbf{Output:}\newline}

 \REQUIRE
    $ $ 
    \begin{itemize}
        \item number of filter taps $m$
        \item filter creation function $C$
        \item model $h$
        \item number of training epochs $T$
        \item learn rate $\eta$
        \item loss function $\mathscr{L}$
        \item training inputs set $\bm{s} = \{\vec{s}_i : \forall i\}$
        \item training labels set $\bm{k} = \{k_i : \forall i\}$
    \end{itemize}

 \ENSURE
    $ $ 
    \begin{itemize}
        \item The adversarial filter $\vec{\delta}_{GAF}$
    \end{itemize}

$ $

\STATE Initialize $\vec{\Delta}_0 \sim \mathcal{N}(0^{1\times l},I^{l \times l})$
\FOR {$t = 1$ to $T$}
    \STATE $\vec{\delta} = C(\vec{\Delta}_{t-1})$
    \STATE $\bm{s}_a = \{\vec{s}_{a_i}: \vec{s}_{a_i}= \vec{s}_i \circledast \vec{\delta} , \forall i\}$
    \STATE $\hat{\bm{k}} = \{\hat{k}_i:\hat{k}_i = h(\vec{s}_{a_i}) , \forall i\}$
    \STATE $\bm{L} = \{L_i: \mathscr{L}(\hat{k_i} , k_i) , \forall i\}$
    \STATE $L = \sum_{\forall i}L_i$
    \STATE $\vec{\Delta}_{t} = \vec{\Delta}_{t-1} + \eta * \nabla_{\vec{\Delta}_{t-1}}L$
\ENDFOR

\RETURN $\vec{\delta}_{GAF} = C(\vec{\Delta}_{T})$

\end{algorithmic}
\end{algorithm}

\subsection{Filter Creation Functions for GAF}
The previous subsection describes how to apply the filter creation function $C$, but does not define it. This subsection proposes possible creation functions. All three proposed creation functions ensure that the power preserving constraint in (\ref{eq:power_preserve}) is satisfied. 

\subsubsection{Unconstrained GAF} One approach to creating a GAF is to treat $\vec{\Delta}$ as an unconstrained filter, and set $\vec{\delta}$ to be a power preserving version of $\vec{\Delta}$. The unconstrained GAF creation function is given by
\begin{equation}
    \vec{\delta} = C_u(\vec{\Delta}) = \frac{\vec{\Delta}}{\||\vec{\Delta}|\|},
\end{equation}
where $\vec{\Delta}$ is complex valued, and $l=m$ for dimensionality. The unconstrained GAF is named so, because this creation function has no constraints on the filter taps besides power preservation, and it does not guarantee an FIR filter with a stable inverse.

\subsubsection{First Tap Constrained GAF} Any minimum phase FIR filter has a stable inverse. Cauchy's argument principle states that if $D(\omega)$ does not wrap around the origin in the complex plane, $\vec{\delta}$ is a minimum phase filter. One way to ensure that $D(\omega)$ does not wrap around the origin it to make sure it's real part is always greater than $0$:
\begin{equation}
    \label{eq:min_phase_constraint}
    Real\{D(\omega)\} > 0 , \forall \omega.
\end{equation}
A simple proposed way to satisfy the constraint in  (\ref{eq:min_phase_constraint}) is to constrain the first filter tap:
\begin{equation}
    \label{eq:first_tap_constraint}
    Real\{\vec{\delta}_0\} > \sum_{i=1}^{m-1}|\vec{\delta}_i|.
\end{equation}
The reasoning behind constraint in (\ref{eq:first_tap_constraint}), is that if all the terms of $D(\omega)$ are lined up to have a phase of $\pi$, a large real vector added to that will keep $D(\omega)$ real. The proposed approach to satisfying (\ref{eq:first_tap_constraint}) is to let the taps of a temporary filter $\vec{h}_{ftc}$ be
\begin{equation}
    \label{eq:min_phase}
    \vec{h}_{ftc,k}=\begin{cases}
            \beta + 1 & k=0\\
            \frac{\vec{\Delta}_{k-1}}{\sum_{i=0}^{m-2}|\vec{\Delta}_i|} & k \in [1,m-1]
        \end{cases},
\end{equation}
where $\beta>0$ is a real constant to ensure stability as opposed to marginal stability, $l=m-1$, and $\vec{\Delta}$ is complex valued. Now the filter creation function for the first tap constrained GAF is
\begin{equation}
    \label{eq:ftc_gaf}
    \delta = C_{ftc}(\vec{\Delta}) = \frac{\vec{h}_{ftc}}{\||\vec{h}_{ftc}|\|}.
\end{equation}

\subsubsection{Root Training GAF} 
\begin{figure}
    \centering
    \includegraphics[trim=0 0 562 500, angle=-90,clip,width=250pt]{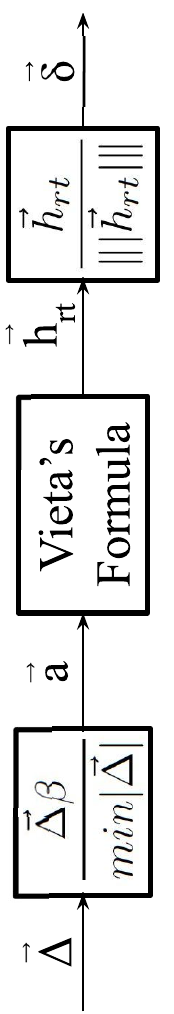}
    \caption{Flowgraph of the filter creation function $C_{rt}(\vec{\Delta})$ for the root training GAF. }
    \label{fig:rtgaf}
\end{figure}

The final proposed approach to creating a GAF with a stable inverse seeks to treat $\vec{\Delta}$ as a vector of complex inverted negative zeros of the filter $\vec{\delta}$. When a filter is inverted the zeros become poles, and for the inverse filter to be stable its poles must be inside the unit circle. Now the gradient ascent process is effectively training the roots of the Z-transform of $\vec{\delta}$, $D(z)$. $\vec{\delta}$ is a power preserving version of an intermediate filter $\vec{h}_{rt}$
\begin{equation}
    \label{eq:Crt}
    \vec{\delta} = C_{rt}(\vec{\Delta}) = \frac{\vec{h}_{rt}}{\||\vec{h}_{rt}|\|},
\end{equation}
and the Z-transform of $\vec{h}_{rt}$ is given by
\begin{equation}
    \label{eq:polynomial}
    H_{rt}(z) = \sum_{i=0}^{m-1}\big(\vec{h}_{rt,i}z^{-i}\big) = \prod_{i=0}^{m-2}\big(z^{-1}+a_i\big).
\end{equation}
Using Vieta's Formula \cite{vieta}, each coefficient of the temporary filter $\vec{h}_{rt}$ can be expressed as a function of $\vec{a}$:
\begin{equation}
    \label{eq:vieta}
    \vec{h}_{rt,m-i} = \sum_{m-2\geq k_{i-2} > k_{i-3} > ... > k_0 \geq 0}\Bigg(\prod_{j=0}^{i-2}\vec{a}_{k_j}\Bigg).
\end{equation}
The elements of $\vec{a}$ are related to the zeros of $\vec{h}_{rt}$ as follows:
\begin{equation}
    H_{rt}(z) = 0,
\end{equation}
\begin{equation}
    \prod_{i=0}^{m-2}\big(z^{-1}+\vec{a}_i\big) = 0,
\end{equation}
\begin{equation}
    z = \frac{-1}{\vec{a}_i}.
\end{equation}
To ensure that the inverse of $\vec{\delta}$ is stable, the elements of $\vec{a}$ need to be constrained by
\begin{equation}
    \Bigg|\frac{-1}{\vec{a}_i}\Bigg| < 1,
\end{equation}
\begin{equation}
    |{\vec{a}_i}| > 1.
\end{equation}
The proposed method of creating $\vec{a}$ from $\vec{\Delta}$ is
\begin{equation}
    \label{eq:stab_zeros}
    \vec{a} = \frac{\vec{\Delta}\beta }{min|\vec{\Delta}|},
\end{equation}
where $\beta>1$ is a real constant to ensure stability as opposed to marginal stability. For the proposed root constrain GAF, $l=m-1$. Combining (\ref{eq:Crt}),(\ref{eq:polynomial}), (\ref{eq:vieta}), and (\ref{eq:stab_zeros}) gives the the full expression for $C_{rt}(\vec{\Delta})$, which can be visually inferred from figure \ref{fig:rtgaf}.

Of the three proposed filter creation functions for the GAF, the root training GAF is the most computationally intense for converting $\vec{\Delta}$ to $\vec{\delta}$. This is because of the recursive nature in expanding the product to a summation in (\ref{eq:polynomial}) using Vieta's formula. The computational complexity of the root training GAF is on the order of $O(n^2)$, whereas the first tap constrained GAF and the unconstrained GAF are on the order of $O(n)$. Note that optimization only happens once offline, and when deploying the filter they all have the same computational complexity.

\subsection{Fast Gradient Filter Method}

The second novel filtering AL algorithm presented in this paper is the fast gradient filter method (FGFM).
This algorithm is similar to the FGM, because it is based on choosing $\vec{\delta}$ to take a small step in the direction of the gradient with respect to loss.
Like the FGM, the FGFM is also creates an input specific $\vec{\delta}$.
Let the adversarial example $\vec{s}_a$ be a perturbation $\vec{r}$ scaled by the the real scalar $\epsilon$ added to an input $\vec{s}$, so the adversarial example generation function
\begin{equation}
    f(\vec{s},\vec{\delta}) = \vec{s}_a = \vec{s} + \epsilon\vec{r},
\end{equation}
where $\vec{r}$ is created by a perturbation generation function $p(\cdot,\cdot)$ with parameter vector $\vec{\delta}$
\begin{equation}
    \vec{r} = p(\vec{s},\vec{\delta}).
\end{equation}
The goal of the FGFM and FGM is to craft adversarial examples that cause some fixed known classifier $h$ to make incorrect predictions. In the eavesdropper scenario, Alice's goal is to ruin Eve's classifier $h_E$. The prediction of $h$ acting on the adversarial example $\vec{s}_a$ is $\hat{k}$, or
\begin{equation}
    \hat{k} = h(\vec{s}_a).
\end{equation}
The metric for the incorrectness of the prediction is characterized by the loss function $\mathscr{L}$ (e.g., cross entropy, or mean squared error), and the true class $k$ associated with the original input $\vec{s}$. The loss on the adversarial input $\vec{s}_a$ is
\begin{equation}
    L(\vec{s}_a) = \mathscr{L}(\hat{k},k).
\end{equation}
The goal of the adversarial attack is to choose $\vec{\delta}$ to make $h$ as incorrect as possible. This is achieved by selecting $\vec{\delta}$ that maximizes the loss $L(\vec{s}_a)$. $L(\vec{s}_a)$ is difficult to maximize directly, so using first order approximation it is estimated as
\begin{equation}
    \label{eq:lin}
    L(\vec{s}_a) = L(\vec{s}+\epsilon \vec{r}) \approx L(\vec{s}) + \epsilon \nabla_{\vec{s}} L(\vec{s})\cdot \vec{r}.
\end{equation}
This approximation leads to the following theorem.

\textbf{\emph{Theorem 1:}} If $\vec{s}_a = \vec{s}+\epsilon \vec{r}$ and $\vec{r}=p(\vec{s},\vec{\delta})$, under the approximation $L(\vec{s}_a) \approx L(\vec{s}) + \epsilon \nabla_{\vec{s}} L(\vec{s})\cdot \vec{r}$, the optimal choice of $\vec{\delta}$ for maximizing $L(\vec{s}_a)$ is:
\begin{equation}
    \vec{\delta} = \big(J_{\vec{\delta}}(\vec{r})\big)^T      \nabla_{\vec{s}}L(\vec{s}).
\end{equation}

\emph{Proof:}
Taking the Jacobian (denoted by $J$) of $L(\vec{s}_a)$ with respect to $\vec{\delta}$ in (\ref{eq:lin}) introduces the need to utilize the property of dot products for Jacobians:
\begin{equation}
\label{eq:jacob_dot}
J(\vec{a}\cdot \vec{b}) = \vec{a}^{T}J(\vec{b}) + \vec{b}^{T}J(\vec{a}),
\end{equation}
which results in:
\begin{equation}
\label{eq:jacob_outcom}
    J_{ \vec{\delta}} ( L ( \vec{s}_a ) ) \approx J_{\vec{\delta}} ( L ( \vec{s}))+ \epsilon \vec{r}^TJ_{\vec{\delta}}(\nabla_{\vec{s}} L(\vec{s}))+\epsilon (\nabla_{\vec{s}} L(\vec{s}))^TJ_{\vec{\delta}}(\vec{r})
.
\end{equation}
$J_{\vec{\delta}}(L(\vec{s}))$ evaluates to a $[1 \times m]$ vector of zeros, because $L(\vec{s})$ is not a function of $\vec{\delta}$. The term $J_{\vec{\delta}}(\nabla_{\vec{s}}L(\vec{s}))$ evaluates to a $[d \times m]$ matrix of zeros, because $\nabla_{\vec{s}}L(\vec{s})$ is not a function of $\vec{\delta}$. This implies that the second term in (\ref{eq:jacob_outcom}) will result in another $[1 \times m]$ vector of zeros. Therefore, the following expression holds for any form of $r(\vec{s},{\vec{\delta}})$:
\begin{equation}
    \label{eq:jacob_outcom_expanded}
    J_{\vec{\delta}}(L(\vec{s}_a)) \approx \epsilon  (\nabla_{\vec{s}}L(\vec{s}))^TJ_{\vec{\delta}}(\vec{r}) 
    = \epsilon (\nabla_{\vec{s}}L(\vec{s}))^T
    \begin{bmatrix}
    \frac{\partial \vec{r}_1}{\partial {\vec{\delta}}_1} &.&.& \frac{\partial \vec{r}_1}{\partial {\vec{\delta}}_m} \\
    .&.&.&.\\
    .&.&.&.\\
    \frac{\partial \vec{r}_d}{\partial {\vec{\delta}}_1} &.&.& \frac{\partial \vec{r}_d}{\partial {\vec{\delta}}_m}
    \end{bmatrix}
.
\end{equation}
It is assumed that the optimization is being done in a linear domain. Therefore the best option for $\vec{\delta}$ is the approximation of $J_{\vec{\delta}}(L(\vec{s}_a))^T$.

In the above proof, a particular structure is assumed for $f(\vec{s},\vec{\delta})$, but $p(\vec{s},\vec{\delta})$ remained unspecified.

In the case of the FGM $\vec{r}$ is denoted as $\vec{r}_{FGM}$ and $\vec{\delta}$ is denoted as $\vec{\delta}_{FGM}$. The FGM is an additive attack, so $\vec{r}_{FGM}$ is the additive perturbation, which implies $p(\vec{s},\vec{\delta}_{FGM})$ is simply $\vec{\delta}_{FGM}$. That is
\begin{equation}
    \vec{r}_{FGM} = \vec{\delta}_{FGM}
    ,
\end{equation}
and $m=d$ for dimensionality. Now the Jacobian is evaluated as
\begin{equation}
    J_{\vec{\delta}_{FGM}}(\vec{r}_{FGM}) = J_{\vec{\delta}_{FGM}}(\vec{\delta}_{FGM}) = I^{d \times d}
    .
\end{equation}
This results in the same perturbation for the FGM algorithm
\begin{equation}
    \vec{\delta}_{FGM} = \epsilon \nabla_{\vec{s}} L(\vec{s})
    .
\end{equation}

For the FGFM, $p(\cdot , \cdot)$ is convolution between vectors of different dimensionality.
\begin{equation}
    \vec{r}_{FGFM} = \vec{s} \circledast \vec{\delta}
    ,
\end{equation}
and $m \neq d$

Before proceeding, convolution between vectors $\circledast$ must be defined. For demonstration purposes, let $m=3$ and $d=5$ (i.e., ${\vec{s}} = [{\vec{s}}_0,{\vec{s}}_1,{\vec{s}}_2,{\vec{s}}_3,{\vec{s}}_4]^T$ and $\vec{\delta} = [\vec{\delta}_0,\vec{\delta}_1,\vec{\delta}_2]^T$). In typical digital convolution (denoted by$\ast$), signals have infinite dimensionality (i.e., $\vec{s} = [...,0,0,\vec{s}_0,\vec{s}_1,\vec{s}_2,\vec{s}_3,\vec{s}_4,0,0,...]^T$). First, normal convolution is between $\vec{s}$ and $\vec{\delta}$ is calculated:

\begin{equation}
    \label{eq:conv_example}
    \vec{s} \ast {\vec{\delta}} =
    \begin{bmatrix}
        .\\
        .\\
        0 \\
        0\\
        {\vec{\delta}}_0\vec{s}_0\\
        {\vec{\delta}}_0\vec{s}_1 + {\vec{\delta}}_1\vec{s}_0 \\
        {\vec{\delta}}_0\vec{s}_2 + {\vec{\delta}}_1\vec{s}_1 + {\vec{\delta}}_2\vec{s}_0 \\
        {\vec{\delta}}_0\vec{s}_3 + {\vec{\delta}}_1\vec{s}_2 + {\vec{\delta}}_2\vec{s}_1 \\
        {\vec{\delta}}_0\vec{s}_4 + {\vec{\delta}}_1\vec{s}_3 + {\vec{\delta}}_2\vec{s}_2 \\
        {\vec{\delta}}_1\vec{s}_4 + {\vec{\delta}}_2\vec{s}_3 \\
        {\vec{\delta}}_2\vec{s}_4 \\
        0\\
        0\\
        .\\
        .\\
    \end{bmatrix}
    ,
\end{equation}

then the center $d$ entries are kept as the result for $\vec{s} \circledast \vec{\delta}$

\begin{equation}
    \label{eq:vect_conv_example}
    \vec{s} \circledast {\vec{\delta}} =  \vec{r} =
    \begin{bmatrix}

        {\vec{\delta}}_0\vec{s}_1 + {\vec{\delta}}_1\vec{s}_0 \\
        {\vec{\delta}}_0\vec{s}_2 + {\vec{\delta}}_1\vec{s}_1 + {\vec{\delta}}_2\vec{s}_0 \\
        {\vec{\delta}}_0\vec{s}_3 + {\vec{\delta}}_1\vec{s}_2 + {\vec{\delta}}_2\vec{s}_1 \\
        {\vec{\delta}}_0\vec{s}_4 + {\vec{\delta}}_1\vec{s}_3 + {\vec{\delta}}_2\vec{s}_2 \\
        {\vec{\delta}}_1\vec{s}_4 + {\vec{\delta}}_2\vec{s}_3 \\

    \end{bmatrix}
    .
\end{equation}

Now the matrix $J_{\vec{\delta}}(\vec{r})$ is given by the Toeplitz structure:

\begin{equation}
    \label{eq:fgfm_Jacobian}
    J_{\vec{\delta}}(\vec{r}) =
    \begin{bmatrix}
        \vec{s}_1 & \vec{s}_0 & 0         \\
        \vec{s}_2 & \vec{s}_1 & \vec{s}_0 \\
        \vec{s}_3 & \vec{s}_2 & \vec{s}_1 \\
        \vec{s}_5 & \vec{s}_3 & \vec{s}_2 \\
        0         & \vec{s}_4 & \vec{s}_3 \\
    \end{bmatrix}
    .
\end{equation}

 Now ${\vec{\delta}}$ can be chosen, which is the filter for creating $\vec{r}_{FGFM}$:

\begin{equation}
    \label{eq:conv_filter}
    {\vec{\delta}}=\epsilon (J_{\vec{\delta}}L(\vec{s}_a))^T \approx \epsilon
            \begin{bmatrix}
                \vec{s}_1 & \vec{s}_2 & \vec{s}_3 & \vec{s}_4 & 0 \\
                \vec{s}_0 & \vec{s}_1 & \vec{s}_2 & \vec{s}_3 & \vec{s}_4\\
                0         & \vec{s}_0 & \vec{s}_1 & \vec{s}_2 & \vec{s}_3\\
            \end{bmatrix}
    (\nabla_{\vec{s}}L(\vec{s}))
    .
\end{equation}

The form of ${\vec{\delta}}$ generalizes for other values of $d$ and $m$. The structure of Toeplitz matrix $J_{\vec{\delta}}(\vec{r})$ depends on how the vector convolution operation $\circledast$ is defined, but its dimensionality is always $[d\times m]$. In the case of the convolution function from the Numpy library \cite{scipy} with the parameter \emph{"padding=same"} , the aforementioned definition of $\circledast$ is the same.

\begin{algorithm}
\caption{Algorithm for creating $\vec{\delta}$ on a specific input using the FGFM.}
\label{alg:fgfm}
\begin{algorithmic}[1]

\renewcommand{\algorithmicrequire}{\textbf{Inputs:}}
\renewcommand{\algorithmicensure}{\textbf{Output:}\newline}

\REQUIRE
$ $ 
\begin{itemize}
    \item number of filter taps $m$
    \item model $h$
    \item loss function $\mathscr{L}$
    \item training input $\vec{s}$
    \item training label $k$
    \item scalar of filter $\epsilon$
\end{itemize}

\ENSURE
$ $ 
\begin{itemize}
    \item The adversarial filter ${\vec{\delta}}_{FGFM}$ designed for the input $\vec{s}$
\end{itemize}

$ $

\STATE Initialize $J_{\vec{\delta}}(\vec{r})^T = 0^{[m \times d]}$

\FOR {$i = 1$ to $m$}
\STATE $J_{\delta}(\vec{r})^T$ row $i$ columns $i-\frac{m}{2}$ to $i-\frac{m}{2}+d-1 = \vec{s}$
\ENDFOR

\STATE $L(\vec{s}) = \mathscr{L}(k,h(\vec{s}))$
\STATE ${\vec{\delta}} = J_{\vec{\delta}}(\vec{r})^T \nabla_{\vec{s}}L(\vec{s})$
\STATE $\vec{v} = 0^{[m \times 1]}$
\STATE $\vec{v}$ column $\frac{m}{2} = 1$
\STATE ${\vec{\delta}}_{FGFM} = \vec{v} + \epsilon {\vec{\delta}}$

\RETURN ${\vec{\delta}}_{FGFM}$

\end{algorithmic}
\end{algorithm}

Now the adversarial filter attack under this approximation is denoted by
\begin{equation}
    \label{eq:bad_filter}
    \vec{s}_a = \epsilon {\vec{\delta}} \circledast \vec{s} +\vec{s}
    .
\end{equation}
Alice ultimately wants to create a filter ${\vec{\delta}}_{FGFM}$ for creating adversarial examples without the need to add the original signal $\vec{s}$. Luckily it is simple to convert the adversarial attack in (\ref{eq:bad_filter}) to an equivalent filter that accomplished the following:
\begin{equation}
    {\vec{\delta}}_{FGFM} \circledast \vec{s} = \epsilon {\vec{\delta}} \circledast \vec{s} + \vec{s}
    .
\end{equation}
${\vec{\delta}}_{FGFM}$ is given by
\begin{equation}
    {\vec{\delta}}_{FGFM} = \epsilon {\vec{\delta}} + \vec{v}
    ,
\end{equation}
where $\vec{v}$ is an $[m \times 1]$ vector of zeros with a $1$ in the center. In the case of $m=3$, $\vec{v}=[0, 1, 0]^T$.

The algorithm for creating an input specific ${\vec{\delta}}_{FGFM}$ is formally described in algorithm \ref{alg:fgfm}. First, the Toeplitz matrix $J_{\vec{\delta}}(\vec{r})^T$ is initialized to a matrix of zeros, and $\vec{s}$ is tiled in a down-right diagonal fashion in each row such that $\vec{s}$ is aligned with $J_{\vec{\delta}}(\vec{r})^T$ in the middle ($m/2$) row. The loss $L(\vec{s})$ is then calculated as the loss function acting on the true label and the model acting on the signal $\mathscr{L}(k,h(\vec{s}))$. Next, the filter for creating an additive perturbation ${\vec{\delta}}$ is computed as $J_{\vec{\delta}}(\vec{r})^T \nabla_{\vec{s}}L(\vec{s})$. Then the vector $\vec{v}$ is set to a $[m \times 1]$ vector of zeros with a $1$ in the center. Finally the FGFM filter ${\vec{\delta}}_{FGFM}$ is computed as $\vec{v}+\epsilon {\vec{\delta}}$ which is returned.

\begin{figure*}[t]
    \includegraphics[trim=25 12 50 50, clip,width=250pt]{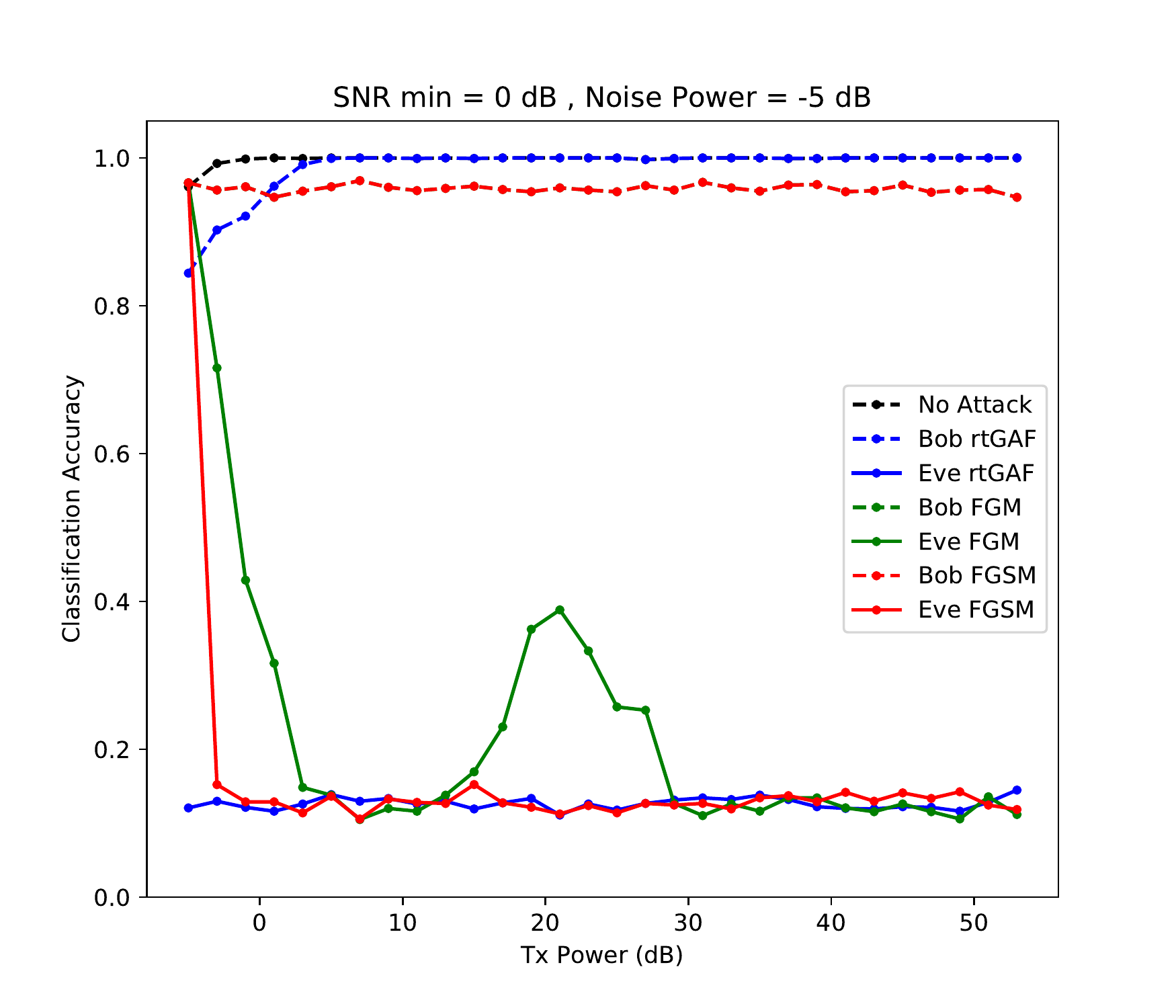}
    \includegraphics[trim=25 12 50 50, clip,width=250pt]{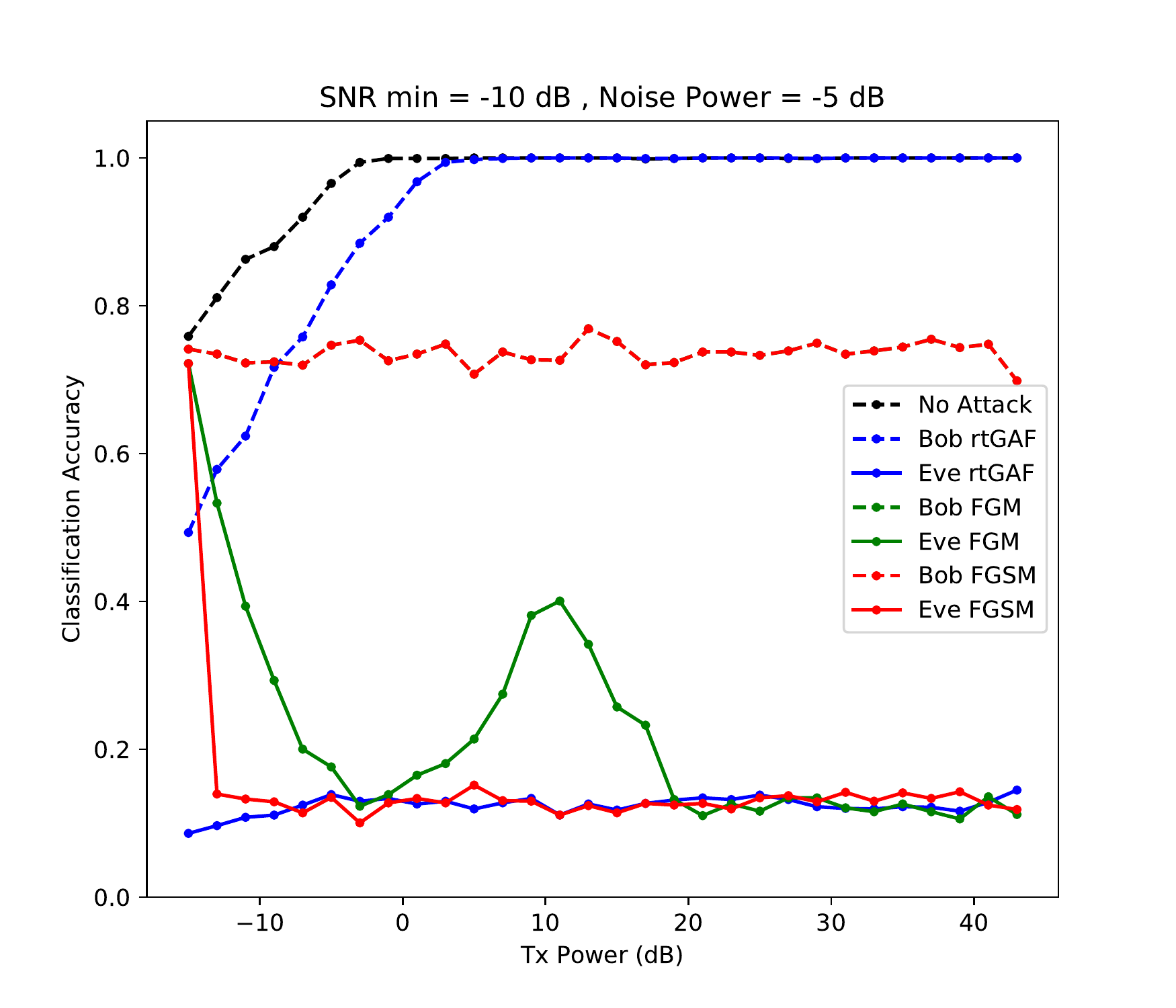}
    \caption{The relationship between Alice's transmit power and Eve/Bob's classification accuracy when different attack are applied by Alice and Bob. The minimum SNR requirement at Bob is 0dB (left), -10dB (right), and the noise power is $5$dB.}
    \label{fig:filter_vs_add}
\end{figure*}

\begin{figure*}[t]
    \includegraphics[trim=25 12 50 50, clip,width=250pt]{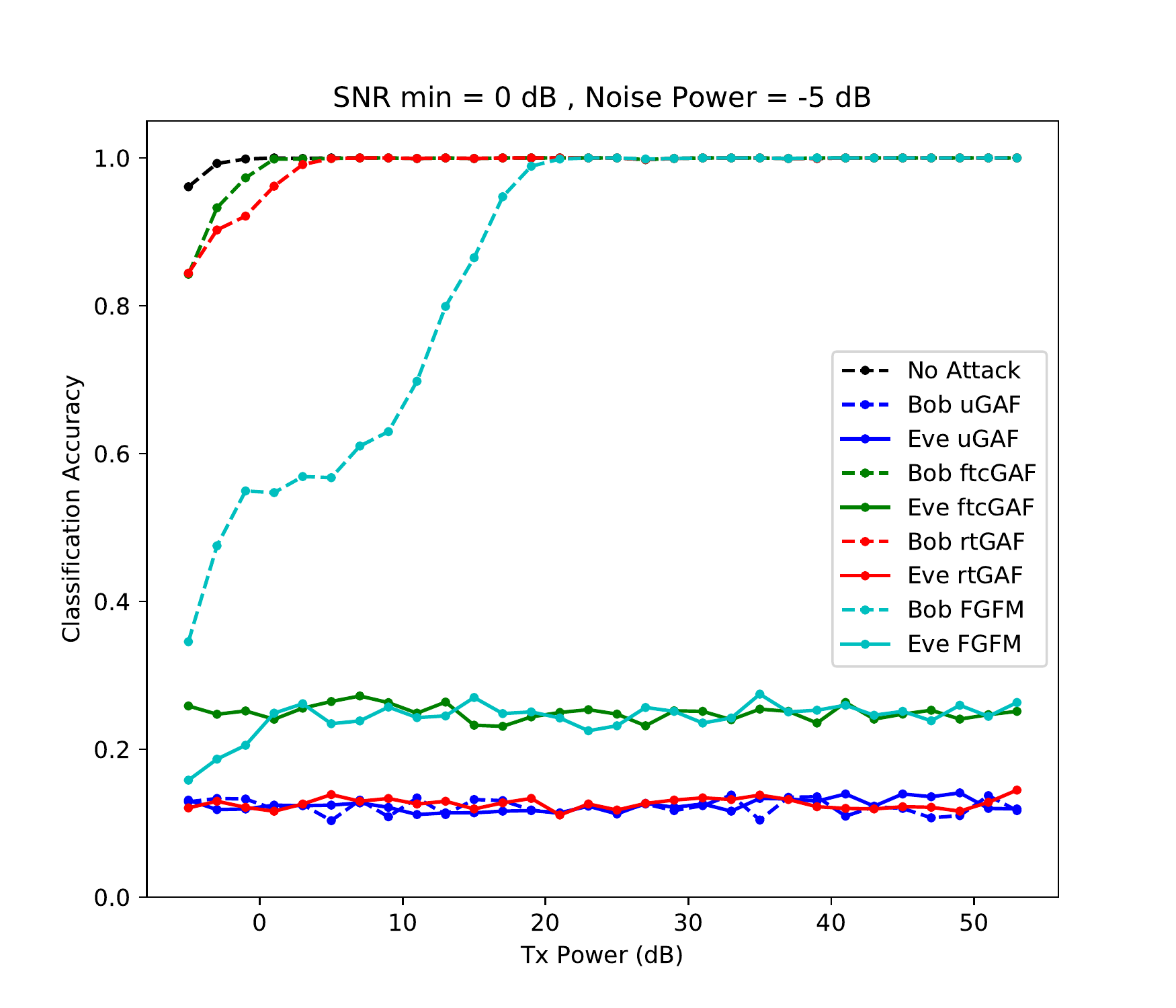}
    \includegraphics[trim=25 12 50 50, clip,width=250pt]{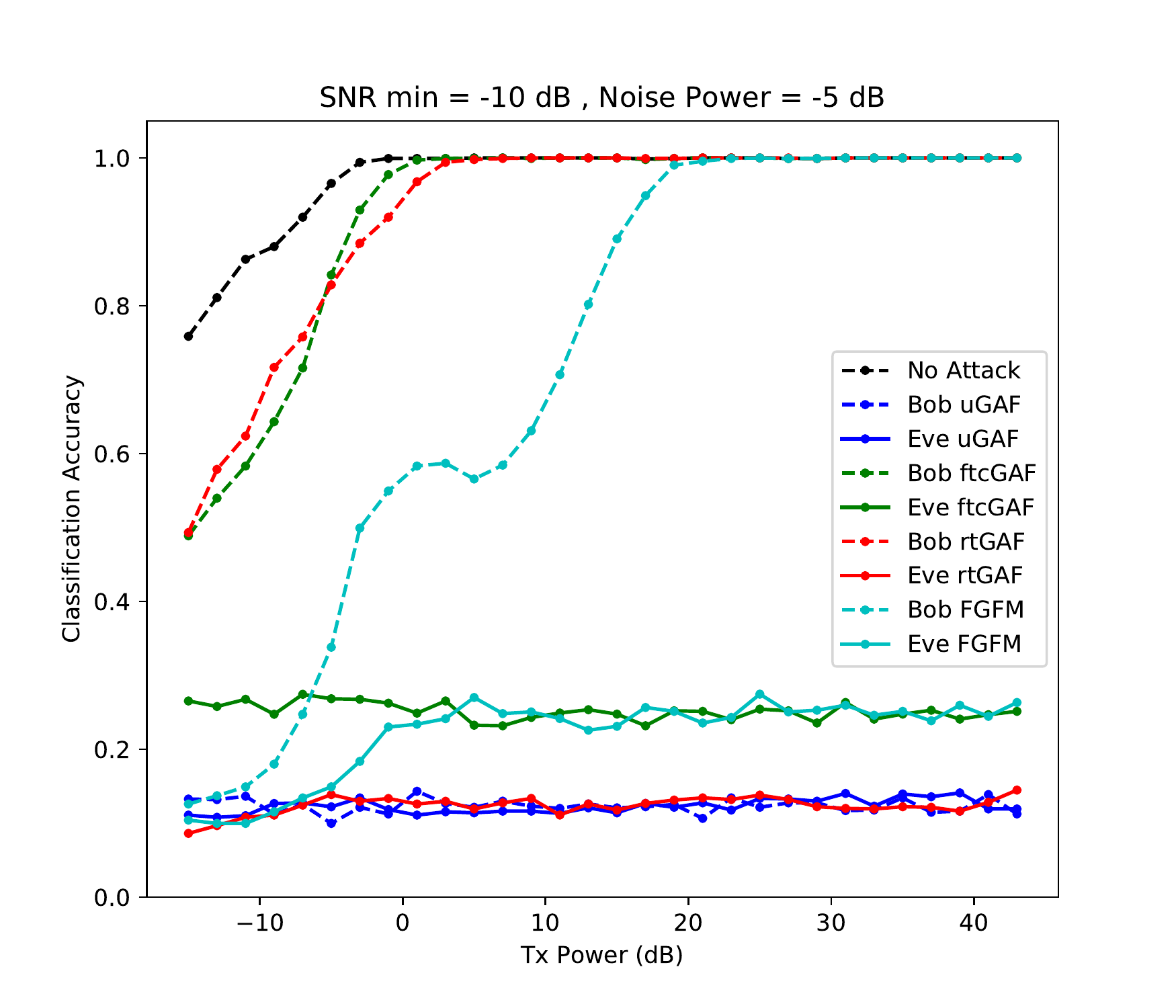}
    \caption{The relationship between Alice's transmit power and Eve/Bob's classification accuracy when different filtering attack are applied by Alice and Bob. The minimum SNR requirement at Bob is 0dB (left), -10dB (right), and the noise power is $5$dB.}
    \label{fig:compare_filters}
\end{figure*}

\begin{figure*}[t]
    \includegraphics[trim=25 12 50 50, clip,width=250pt]{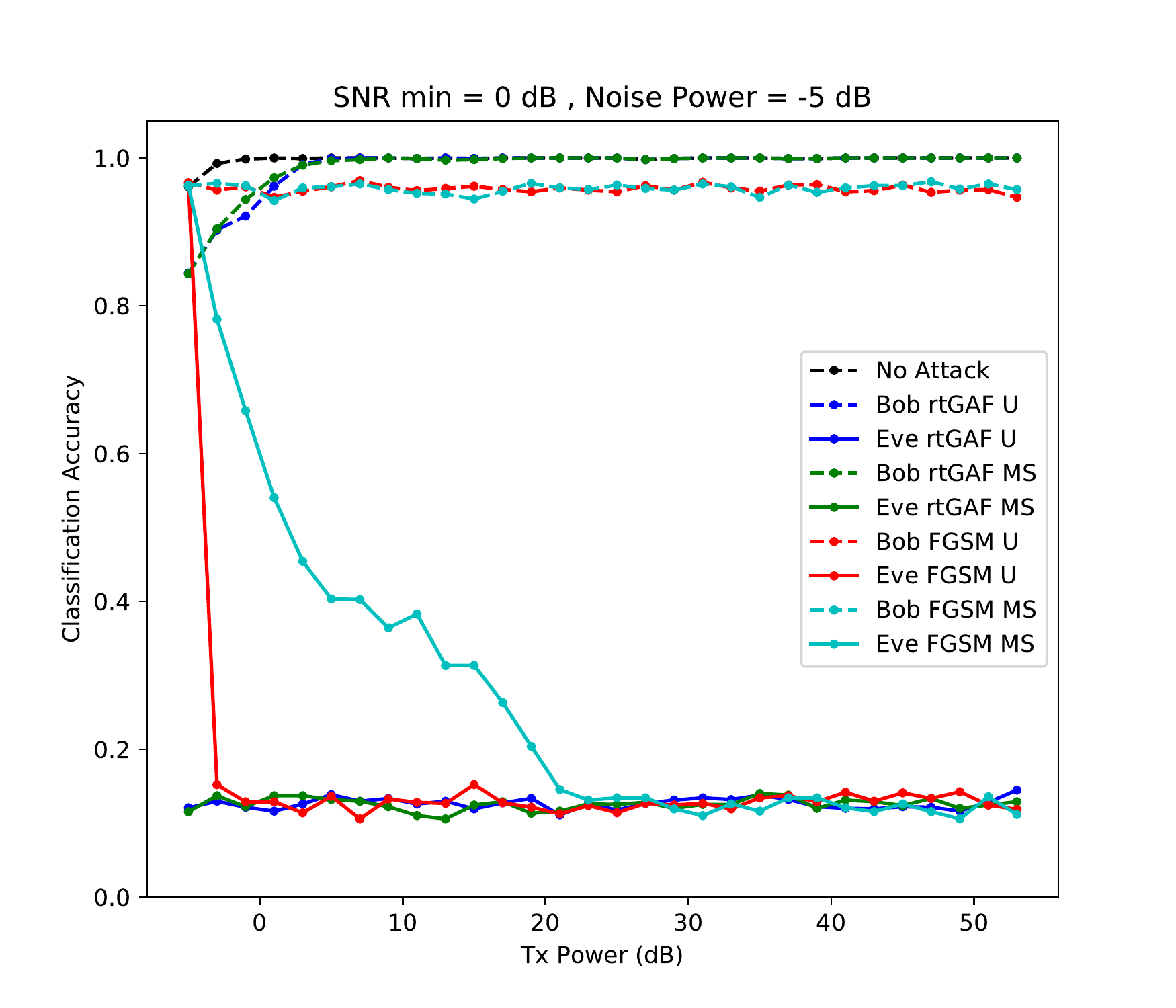}
    \includegraphics[trim=25 12 50 50, clip,width=250pt]{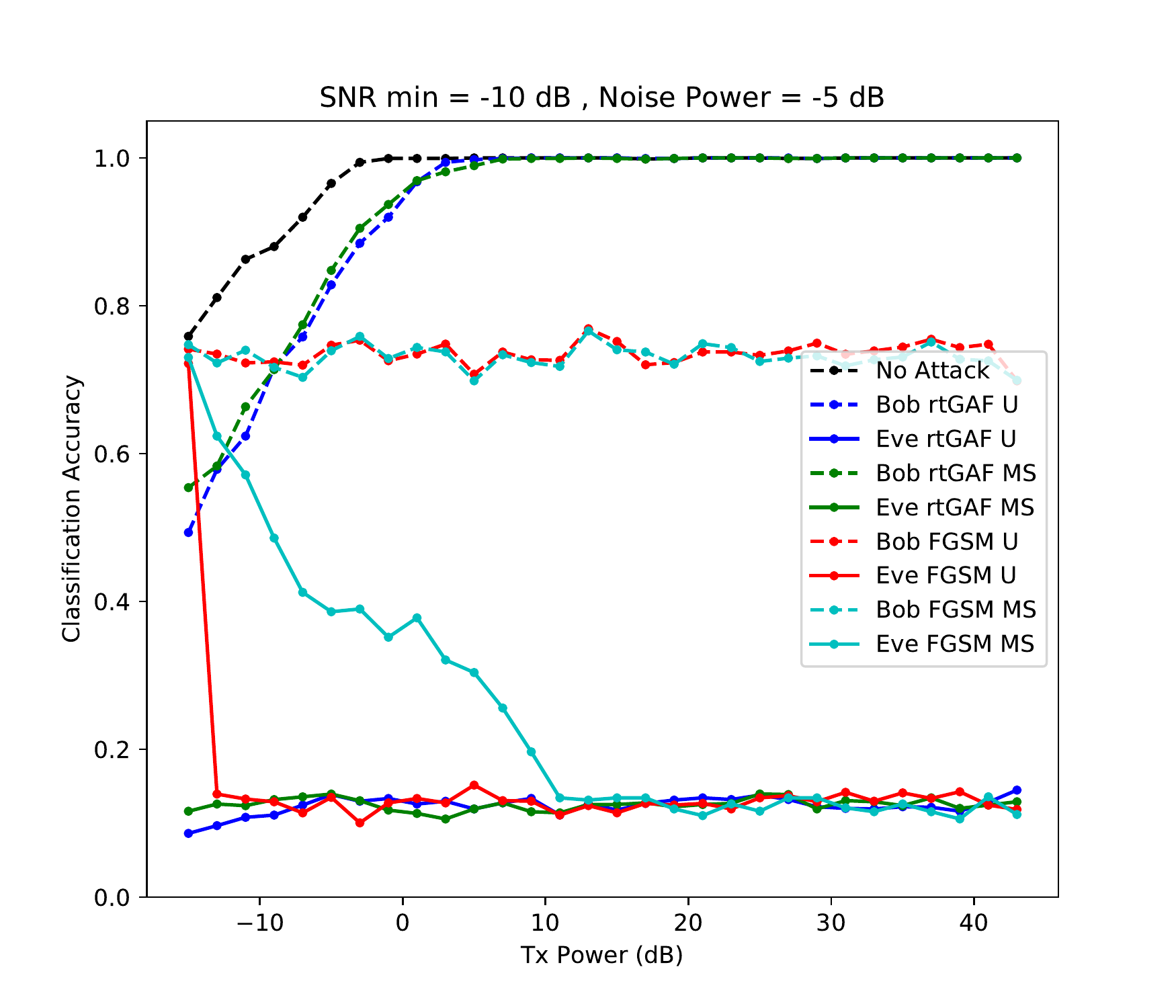}
    \caption{The relationship between Alice's transmit power and Eve/Bob's classification accuracy when root training GAF and the FGSM applied by Alice and Bob as modulation specific attacks (MS) and universal attacks (U). The minimum SNR requirement at Bob is 0dB (left), -10dB (right), and the noise power is $5$dB.}
    \label{fig:ms_vs_u}
\end{figure*}

\begin{figure}
    \centering \includegraphics[trim=25 12 50 50, clip,width=250pt]{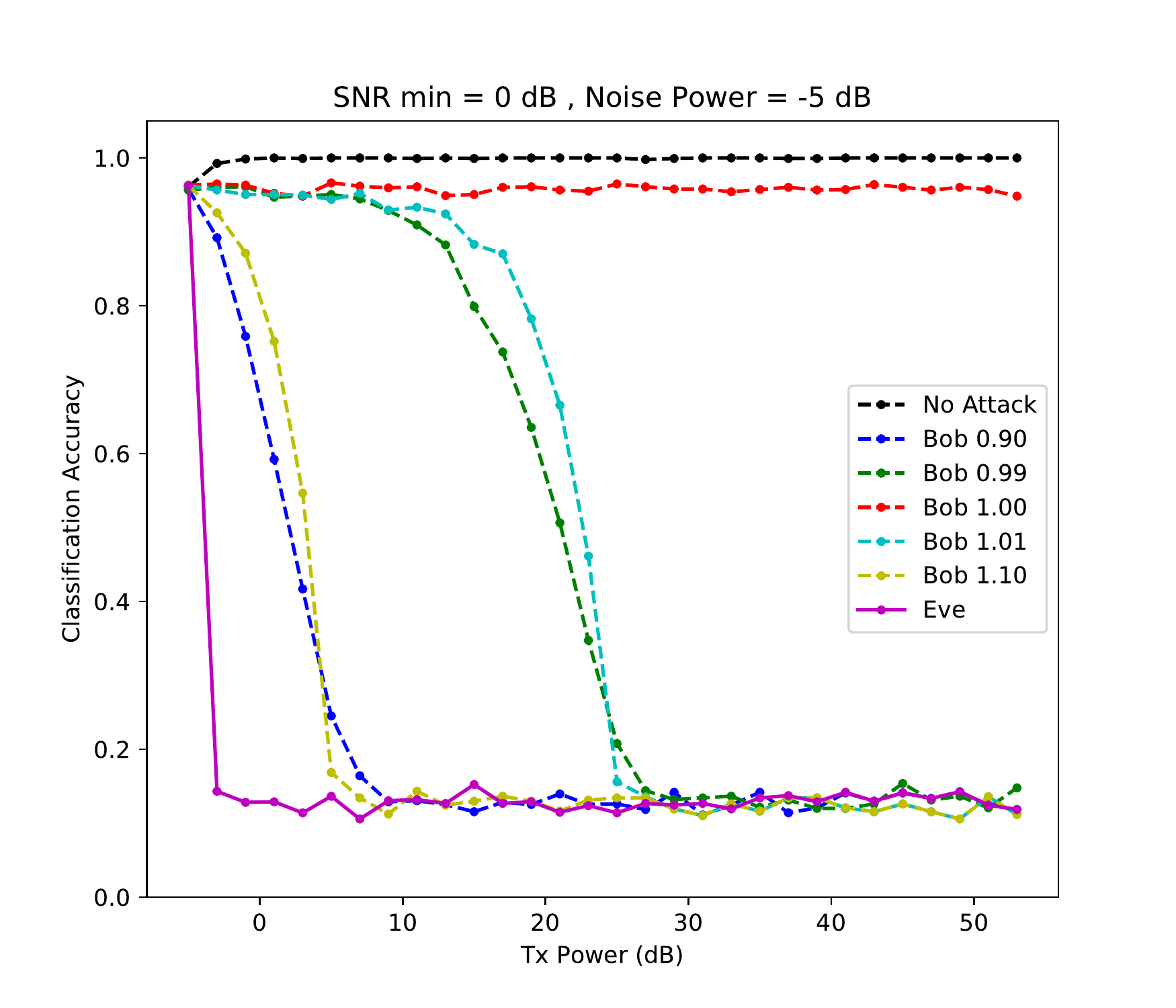}
    \caption{The relationship between Alice's transmit power and Eve/Bob's classification accuracy when  noise power is $5$dB and the minimum SNR requirement at Bob is 0dB. The attack used is the FGSM, and the numbers in the labels indicate the ratio between Bob's estimated channel attenuation and the true attenuation $\hat{\alpha}/\alpha$.}
    \label{fig:alpha_anal}
\end{figure}

Recall that ${\vec{\delta}}_{FGFM}$ is specifically to designed for a particular input $\vec{s}$. As previously stated in section \ref{sec:results} a universal attack is needed for Alice to effectively communicate with Bob. One can simply use existing algorithms to aggregate many filters generated by the FGFM into a universal filter, such as the PCA based approach presented in \cite{sadeghi_original}.

It is important to note that the FGFM may not be a minimum phase filter, hence Bob may not be able to use a stable inversion to get Alice's signal back.

\section{Experiments}
\label{sec:experiments}
All experiments were conducted purely in simulation using Python \cite{python}. The TensorFlow \cite{tensorflow} and Keras \cite{keras} libraries were used as the platform for implementing DL algorithms with graphics processing unit (GPU) acceleration. Many useful functions, such as convolution, from the SciPy \cite{scipy} library, were used.
The machine-learning algorithm analyzed in these simulations is a deep convolutional neural network with inputs that have normalized power and mean.
The aforementioned DL architecture was used for both Bob and Eve.
Only white-box attacks were considered, however it has been shown in many AL works that AL attacks are transferable between DL models.
The channel model used is the attenuation and AWGN described in section \ref{sec:sys_model} with trivial attenuation $\alpha=1$. Adjusting attenuation with this system model is analogous to adjusting transmit power and noise power.
To implement the FGFM, FGM, and FGSM as universal or modulation specific attacks, the PCA based algorithm for creating UAP's from Sadeghi et al was used \cite{sadeghi_original}.
All filters in these experiments have 5 taps.
Classification accuracy in these experiments is given by averaging the probability of correct classification over all classes.



Figure \ref{fig:filter_vs_add} show experiment results when the minimum SNR requirement at Bob is $0$ and $5$ dB respectively. Eve's classification accuracy, as well as Bob's, is plotted against Alice's available transmit power. In this experiment it is assumed that Bob can estimate the attenuation of the channel $\alpha$ perfectly, so the plot shows that Bob's classification accuracy does not drop in additive attacks.
The lowest Tx power shown in the plots indicates that $P_T$ is just enough to satisfy the SNR requirement at Bob. This means that additive attacks such as the FGSM or FGM cannot have any power allocated to them (i.e $P_T = P_{\vec{s}} + P_{\vec{\delta}}$ and $P_{\vec{\delta}} = 0$). Eve's classification accuracy is equal to Bob's classification accuracy when $P_T$ is very low because of this.
The FGSM has a steeper dropoff in Eve's classification accuracy than the FGM because the FGSM is the optimal solution to minimizing perturbation power (measured by the infinite norm, not L2 norm) such that classification is incorrect. The FGM was not derived to minimize perturbation power, so it does not drop Eve's classification accuracy as steep.
However, when $P_T$ is very low the root training GAF (labeled as rtGAF) reduces Eve's classification accuracy, because there is no need to dedicate transmit power for the attack.
Bob's classification accuracy remains constant under the additive attacks because Alice is allocating as much transmit power to the perturbation as possible. Bob's classification accuracy raises with Alice's transmit power when filter attacks are used because Alice can allocate more power to her signal and Bob can almost perfectly recover the un-attacked signal.

Figure \ref{fig:compare_filters} shows experimental results comparing Eve and Bob's classification accuracy when Alice employs different filtering attacks proposed in this paper.
The FGFM produces a very effective filter at fooling Eve, however, it's inverse is unstable so Bob's classification accuracy suffers as well. The unconstrained GAF (labeled as uGAF) was the most effective at fooling Eve, due to being able to optimize filter taps unconstrained, but its inverse is even more unstable than the FGFM which causes Bob's classification accuracy to be worse.
The first tap constrained GAF (labeled as ftcGAF) has a stable inverse, so Bob's classification accuracy is good. However, the first tap constrained GAF retains a large portion of Alice's original because of the large real-valued first tap, so Eve is not fooled effectively. $\beta$ for the first tap constrained GAF was $0.9$ for this experiment.
The root training GAF (labeled as rtGAF) also has a stable inverse, and it is significantly less constrained than the first tap constrained GAF, so it is intuitive that Eve's classification accuracy is very low with this filter. $\beta$ for the root training GAF was $0.5$ for this experiment.

Figure \ref{fig:ms_vs_u} shows experimental results comparing Eve and Bob's classification accuracy when Alice employs a universal version and modulation specific version of the root training GAF and the FGSM. It is clear from this experiment that the modulation specific version of the root training GAF is more effective at fooling eve than the universal version. This is because the GAF is trained to be near-optimal for every modulation format in the modulation specific attack case, whereas in the universal attack case the GAF must compromise between all modulation formats during optimization. It is interesting to see that the FGSM is not as effective when modulation specific as opposed to universal. This is likely because of the ineffectiveness of the algorithm used to create a universal adversarial perturbation from the many individual perturbations.

To further emphasize the importance of Bob estimating the channel attenuation coefficient $\alpha$ accurately, figure \ref{fig:alpha_anal} shows Bob's classification accuracy when his estimate $\hat{\alpha}$ is not perfect. In this experiment, only the FGSM is analyzed. Each curve represents Bob's classification accuracy on his recovered signals as indicated by (\ref{eq:alpha_hat}). The number used to label each curve is the ratio between Bob's estimate and the true channel attenuation coefficient $\hat{\alpha}/\alpha$. The perfect estimate is when $\hat{\alpha}/\alpha=1$ where Bob's classification accuracy is highest. As $\hat{\alpha}/\alpha$ deviates further from $1$, Bob's classification accuracy drops detrimentally. An imperfect estimate of the channel's attenuation means that when Alice has more transmit power dedicated to the perturbation ${\vec{\delta}}_+$, the leftover perturbation from Bob's signal recovery becomes larger. Eve's classification accuracy is not affected by Bob's inability to estimate $\alpha$

\section{Conclusion}
\label{sec:conc}
This paper examines point to point communication system with an eavesdropper (Eve), where the transmitter (Alice) has finite transmission power, and the intended receiver (Bob) has a minimum SNR requirement that must be met for reliable communication.
The channel model explored in this work is an AWGN channel model with attenuation of the transmitted signal, and the channel effects are assumed to be the same for Bob and Eve.
Secrecy is measured by making Eve's classification accuracy low, but keeping Bob's classification accuracy high.
Alice uses an adversarial attack on her transmitted signal to lower Eve's classification accuracy and shares a key with Bob so that he can undo the adversarial attack.
Eve does not know the key, so she cannot undo the adversarial attack.
This paper concludes that additive adversarial attacks fall short in this application, because they require sacrificing transmit power, and can be difficult to undo for Bob.
To undo an additive perturbation, Bob must synchronize the subtraction of the perturbation with Alice's repeated addition of the perturbation. In addition to synchronization, Bob must also predict the attenuation coefficient of the channel, which if not done perfectly will leave remnants of the adversarial perturbation which is detrimental to classification accuracy.
If Alice uses filters to generate adversarial examples instead of additive perturbations, there is no need to sacrifice transmit power and it is simpler to undo the adversarial attack at Bob.
Bob needs only to use the inverse to the filter used at Alice to undo the adversarial attack.

This paper also presents two novel filter-based AL algorithms to generate adversarial examples.
The first of which is an optimization approach called the gradient ascent filter (GAF) which is similar to training a neural network to minimize loss, but instead the filter is being trained to maximize loss of a fixed machine-learning model. This paper proposes three methods of creating a GAF, two of which have a stable inverse which is necessary for Bob to decode information.
The second presented AL filtering algorithm is similar to the fast gradient method (FGM), called the fast gradient filter method (FGFM). This algorithm, like the FGM, is the solution to a first-order approximation of loss when an adversarial example is predicted by a fixed machine-learning model. Unlike the two stable GAF algorithms, the FGFM has an unstable inverse which makes it infeasible for Bob to use.
In simulations with a convolutional neural network, the root training GAF was the most effective AL algorithm in this system model where transmit power is limited.

\bibliographystyle{unsrt}
\bibliography{ms}  






\end{document}